# Fe-FeH Eutectic Melting Curve and the Estimates of Earth's Core Temperature and Composition

Shuhei Mita[1], Shoh Tagawa[1,2], Kei Hirose[1,3], and Nagi Ikuta[1]

[1]Department of Earth and Planetary Science, The University of Tokyo, Tokyo, Japan
[2]Now at Center for Innovative Higher Education, Chiba University, Chiba, Japan
[3]Earth-Life Science Institute, Tokyo Institute of Technology, Tokyo, Japan

Correspondence to:
S. Mita
mita-s@eps.s.u-tokyo.ac.jp

**Key Points:**

- We determined the eutectic melting temperature in the Fe-FeH system between 52 and 175 GPa in a DAC.
- The d*T*/d*P* slope of the Fe-FeH eutectic melting curve is comparable to those of the melting curves of Fe and FeH endmembers.
- We explored a set of possible outer core composition and ICB temperature, which explains the observed outer core density.

**Abstract**

Fe and FeH form a binary eutectic system above ~40 GPa. Here we performed melting experiments in a laser-heated diamond-anvil cell (DAC) and obtained the Fe-FeH eutectic melting curve between 52 and 175 GPa. Its extrapolation shows the eutectic temperature to be 4700 K at the inner core boundary (ICB), which is lower than that in Fe-FeSi but is higher than those in the Fe-S, Fe-O, and Fe-C systems. In addition, its d*T*/d*P* slope is comparable to those of the melting curves of Fe and FeH endmembers, suggesting that the eutectic liquid composition changes little with increasing pressure and is about $FeH_{0.6}$ at the ICB pressure. We also estimated the effect of each light element on depressing the liquidus temperature at 330 GPa based on a combination of binary eutectic temperature and composition and found that the effect is large for C and S, moderate for H and O, and small for Si when considering the amount of each element that reduces a certain percentage of a liquid iron density. Furthermore, we searched for a set of possible outer core liquid composition and ICB temperature (the liquidus temperature of the former at 330 GPa should match the latter), which explains the outer core density deficit that depends on core temperature. The results demonstrate that relatively low core temperatures, lower than the solidus temperature of a pyrolitic lowermost mantle at the core-mantle boundary (CMB), are possible when the core is poor in Si.

**Plain Language Summary**

We examined the eutectic melting in the Fe-FeH system to 175 GPa. The extrapolation of the present data shows the temperature and composition of the eutectic point to be 4700 K and $FeH_{0.6}$, respectively, at 330 GPa. Based on these results and literature data on other Fe alloy systems, we compared the effect of reducing the liquidus temperature of Fe and found that when considering the amount of each light element that explains a given core density deficit, the effect is large for C and S and moderate for H and O. In addition, we explored a set of possible outer core liquid composition and ICB temperature, which accounts for the observed outer core density (the liquidus temperature of a liquid that explains the outer core density under a particular ICB temperature should match such ICB temperature). The results demonstrate that as far as the core is depleted in Si, the ICB and overall core temperatures can be relatively low, less than ~3500 K at the CMB (~4800 K or less at the ICB), which is lower than the solidus temperature of a pyrolitic mantle material and thus avoids extensive melting at the bottom of the mantle.

## 1. Introduction

Hydrogen has been considered one of the important light elements in the Earth's core, but the Fe-FeH phase diagram is little known yet. Since the least amount of hydrogen is soluble into iron at 1 bar, the Fe-FeH phase diagram has been examined under high pressures (>~2 GPa) (Yagi & Hishinuma, 1995; Ikuta et al., 2019). It is necessary to determine the hydrogen content in iron under high pressure, otherwise hydrogen escapes from the Fe lattice upon pressure release (Iizuka-Oku et al., 2017; Tagawa et al., 2021). While Fukai (1992) speculated a continuous solid solution between Fe and FeH above 100 GPa, more recent experimental studies demonstrated that Fe and FeH form a binary eutectic system above ~40 GPa (Hikosaka et al., 2022; Oka et al., 2022; Tagawa, Helffrich et al., 2022), where non-magnetic stoichiometric FeH melts congruently with a temperature maximum. The earlier experiments on Fe-FeH by Oka et al. (2022) reported its melting temperature to be 1900 K at 45 GPa. The Fe-FeH eutectic melting temperature has not been examined at higher pressures, while the melting curve of stoichiometric FeH was determined to 152 GPa (Tagawa, Helffrich et al., 2022). The eutectic melting curves, as well as eutectic liquid compositions, have been previously reported in other binary Fe alloy systems such as Fe-$Fe_3S$ to 254 GPa (Mori et al., 2017; Thompson et al., 2022), Fe-FeSi to 101 GPa (Fischer et al., 2013), Fe-FeO to 204 GPa (Oka et al., 2019; Morard et al., 2017), and Fe-$Fe_3C$ to 255 GPa (Mashino et al., 2019).

In this study, we determined the eutectic melting temperature in the Fe-FeH system by heating a sample consisting of both the H-poor hexagonal close-packed (hcp) and H-rich face-centered cubic (fcc) phases between 52 and 175 GPa in a laser-heated DAC. We employed three independent melting criteria, which gave eutectic melting temperatures at each pressure that are reasonably consistent among them. The eutectic liquid composition was also constrained from the compositions of coexisting liquid and solid (liquidus phase) in this study. The Fe-FeH eutectic temperature and composition extrapolated to 330 GPa are important to explore the outer core liquid composition along with the ICB temperature, which corresponds to the liquidus temperature of an outer core liquid. Combining with earlier experimental data on other binary Fe alloy systems, we searched for a set of possible outer core composition and ICB temperature, which explains the observed outer core density.

## 2. Experimental Methods

High-pressure and -temperature (*P-T*) experiments were performed in a laser-heated DAC. We used diamond anvils with beveled 90, 120, and flat 300 µm culet size, depending on a target pressure. A rhenium gasket was pre-indented to 28–45 µm in thickness. Diamond anvils were coated with Ti by sputtering in order to prevent hydrogen penetration into diamonds (Ohta et al., 2015). A pure Fe foil (>99.999% purity, Toho Zinc) with ~10 μm thickness was loaded into a hole at the center of a pre-indented Re gasket between the KCl (or single crystal $Al_2O_3$ in runs #4 and #6) thermal insulation layers, leaving large room for hydrogen. Only in run #5, a 1 μm thick $ZrO_2$ layer was also sputtered as an additional thermal insulator. A whole DAC was dried in a vacuum oven after the sample was loaded. Supercritical hydrogen fluid was then introduced into a sample chamber by using a high-pressure gas apparatus (PRETECH Co., Ltd) at SPring-8 except for runs #4 and #6. A high-pressure vessel and a gas line were vacuumed for 5 min and then purged four times by $H_2$ gas before loading it into the DAC. We confirmed the existence of $H_2$ molecules by Raman spectra at ~10 GPa. In runs #4 and #6, hydrogen was loaded cryogenically at temperatures below 20 K in a liquid hydrogen-introducing system (Tagawa et al., 2016; Tagawa, Gomi et al., 2022). In these two experiments, double hcp (dhcp) $FeH_x$ ($x$~1.0) was synthesized at 9 GPa by thermal annealing with laser, which was confirmed by X-ray diffraction (XRD) measurements at SPring-8 (see below).

Melting experiments were conducted with *in-situ* high *P-T* XRD measurements at BL10XU, SPring-8. The incident X-ray beam was monochromatized to ~30 keV and focused to 6 μm in diameter. XRD pattens were collected on a flat panel X-ray detector (Varex Imaging, XRD1611CP3 and Perkin Elmer, XRD0822 CP23). Sample was heated from both sides with a couple of 100 W single-mode Yb fiber lasers using the beam shaping optics that makes a laser beam with a flat-top energy distribution. The laser spot size was adjusted to 20–25 μm. One-dimensional radial temperature profile was collected by a spectro-radiometric method using thermal radiation spectra in the range of 600–800 nm across a heated spot (Hirao et al., 2020). Experimental temperature was the average in a 6 μm area from which XRD data were obtained. Pressure was determined from the lattice volume of KCl (Tateno et al., 2019) in runs #1–3 and #5. To estimate the pressure at high temperature, we consider the temperature of the KCl layer to be $T_{\mathrm{KCl}} = (3 \times T_{\mathrm{sample}} + 300) / 4$ by following Campbell et al. (2009). In runs #4 and #6, we obtained the pressure at 300 K from the unit-cell volume of post-perovskite-type $Al_2O_3$ (Caracas & Cohen, 2005) and added the contribution of thermal pressure, +5% per every 1000 K increase (Hirose et al., 2019). Pressure uncertainty may be ±10% for these two runs, while it may be ±5% for other runs with KCl (Hikosaka et al., 2022). Temperature error would be ±5% (Mori et al., 2017).

After compression to a target pressure under room temperature, we heated the sample to about 1000 K and confirmed the coexistence of hydrogen-poor hcp and hydrogen-rich fcc phases in each experiment except for run #6. Subsequently we melted the hcp + fcc sample to determine the eutectic melting temperature between Fe and FeH (Table 1). We repeated the heating-quenching cycles with increasing the laser power output with a 0.5 to 5 W step. Melting was detected based on three different melting criteria (see below).

In addition, we also determined hydrogen concentrations, $x$ in $FeH_x$, in liquid and the coexisting liquidus phase (solid) as well as those in the hcp and fcc phases below eutectic temperature (Table 2), based on their lattice volumes observed after quenching temperature. The hydrogen content was calculated as;

$$x = \frac{V_{\mathrm{FeH}_x} - V_{\mathrm{Fe}}}{\Delta V_{\mathrm{H}}} \tag{1}$$

where $V_{FeH_x}$ and $V_{Fe}$ are the volumes per chemical formula (half and quarter of the unit-cell volumes of hcp and fcc, respectively). We employed the hcp and fcc $V_{Fe}$ from Dewaele et al. (2006) and Dorogokupets et al. (2017), respectively. $\Delta V_H$ is the expansion of the lattice volume per iron atom by incorporating a hydrogen atom and has been estimated by Tagawa, Gomi et al. (2022). The uncertainty in $x$ may be ±7% due to the error in $\Delta V_H$. The hydrogen content in quench crystals formed from liquid upon temperature quench should represent that of liquid (Tagawa et al., 2021) since Fe-H liquids were quenched fully into quench crystals as observed in scanning ion microscope (SIM) images of sample cross sections obtained using a focused-iron beam (FIB, Thermo Fisher Scientific, Helios 5 UC). Compared to the liquid and liquidus phase formed upon melting, both the hcp and fcc phases found below solidus temperature exhibited relatively large variations in the lattice volume and thus hydrogen concentration. We therefore show the range of variations in these subsolidus phases in Table 2.

## 3. Results

### 3.1. Eutectic Melting in the Fe-FeH System

We performed a total of six separate experiments in a pressure range from 52 to 175 GPa (Tables 1, 2). In each experiment except run #6, the H-poor hcp and H-rich fcc phases coexisted before increasing temperature to above melting temperature. We employed three criteria for melting in this study; 1) change in the relation between the laser power output and the sample temperature (temperature plateau, Figure 1), 2) appearance of quench crystals upon quenching temperature, and 3) disappearance of one of subsolidus phases.

In run #1 conducted at 52 GPa, the broad XRD peaks showed the coexistence of hcp $FeH_{\sim 0.3}$ and fcc $FeH_{\sim 0.9}$ before heating (Figures 2a). When heated to 1610 K, the hydrogen content in the hcp phase increased to $x$ = 0.51–0.67, while that in the fcc phase was $x$ = 0.86–0.93 (Figures 2b, c). The XRD patterns remained similar by heating up to 1850 K. On the other hand, when the temperature was increased to 2070 K, the hcp phase disappeared, while the fcc phase remained (Figure 2d). Upon quenching temperature, fcc $FeH_{0.71}$ formed from liquid, coexisting with fcc $FeH_{0.87}$ that was already present during heating (Figure 2e). Furthermore, while we introduced higher laser power output, temperature stayed at 2070–2080 K, showing a temperature plateau (Figure 1). These observations—the loss of one of solid phases, the appearance of quench crystals, and the temperature plateau—consistently indicate the eutectic melting temperature to be 2070–2080 K or at least between 1850 and 2080 K at 52 GPa. Similarly in run #2, eutectic melting was confirmed by these three melting criteria at 62 GPa (Figure 1).

In run #3, the sample was compressed under 300 K to 93 GPa. The XRD pattern exhibited broad peaks from fcc $FeH_{\sim 1.0}$ in addition to hcp Fe. When the sample was heated to 1720 K, the fcc peak intensity became stronger than that of hcp. The fcc phase decreased its hydrogen concentration to $x$ = 0.8–0.9, while the hcp phase was hydrogenated with $x$ = 0–0.4. Furthermore, upon heating to 2710 K at 116 GPa, the fcc and hcp showed $x$ = 0.79–0.96 and 0.49–0.61, respectively (Figure 3a). No new peaks appeared upon quenching temperature to 300 K (Figure 3b). On the other hand, when we increased the sample temperature to 3060 K at 118 GPa, the hcp phase was lost, while the fcc ($x$ = 0.93) was still observed (Figure 3c). After temperature quench, the new fcc peaks appeared with a lesser amount of hydrogen ($x = 0.74$), which should have derived from liquid (Figure 3d). The new peaks from liquid were relatively weak, suggesting that the bulk hydrogen content in the area involved in melting was close to that of the liquidus fcc phase. These indicate that the eutectic temperature is between 2710 K and 3060 K at 116–118 GPa. Since we did not further increase the laser power output, a temperature plateau was not examined in this experiment.

In runs #4 and #6, dhcp $FeH_{\sim 1.0}$ was synthesized by thermal annealing at ~9 GPa. Upon further compression at 300 K, it fully changed into the fcc structure. In the former run, fcc $FeH_{\sim 0.8}$ was found at ~140 GPa before reheating. We observed the coexistence of hcp $FeH_{0.22–0.46}$ and fcc $FeH_{0.96}$ up to 2590 K at 148 GPa. When the sample temperature was increased to 3150 K at 151 GPa and then gradually decreased, several hcp spot peaks in the 2D XRD image became stretched (Figure 4). It may be attributed to an overgrowth of the liquidus phase of hcp upon decreasing temperature, indicating partial melting, although the fcc phase still remained in the XRD pattern. In run #6 performed at ~174 GPa, no extra XRD peaks were observed upon quenching temperature from 3220 K. When it was reheated to 3400 K, we found additional peaks from the quench crystals of fcc $FeH_{0.67}$ coexisting with fcc $FeH_{1.05}$ in the XRD pattern collected at 300 K after quenching. While the hcp phase was not confirmed, this liquid composition is likely to be close to a eutectic composition ($x = \sim 0.6$, Figure 5) (see below), suggesting that melting occurred at temperature close to the eutectic temperature. Finally, we obtained no evidence for melting up to 3200 K at 159 GPa in run #5, which gives the lower bound for the eutectic temperature.

Based on these observations described above, we obtained the Fe-FeH eutectic melting curve (Figure 6). The lower and upper bounds for melting temperature at each pressure were fitted by Simon-Glatzel equation (Simon & Glatzel, 1929) (Table 1);

$$T = T_0 \times [(P - P_0)/a + 1]^{1/c} \quad (1)$$

in which we fixed $P_0 = 45$ GPa since Fe-FeH is not a eutectic system at lower pressures (Tagawa, Helffrich et al., 2022). The fitting provided $T_0 = 1830$ K, a = 64, and c = 1.8. The eutectic melting curve obtained is consistent with all the data collected in this study and Oka et al. (2022). The present Fe-FeH eutectic melting curve is higher than the eutectic temperature between FeH and $FeH_2$ previously reported by Hirose et al. (2019) (Figure 6).

## 4. Discussion

### 4.1. Depression of Liquidus Temperature of Iron by Incorporating Light Elements

The extrapolation of the melting curve shows the Fe-FeH eutectic melting temperature to be 4700 K at the ICB (Figure 7). The eutectic melting curves have been already reported for other binary Fe alloy systems; Fe-$Fe_3S$ (Mori et al., 2017; Thompson et al., 2022), Fe-FeSi (Fischer et al., 2013), Fe-FeO (Oka et al., 2019; Morard et al., 2017), and Fe-$Fe_3C$ (Mashino et al., 2019). Figure 7 compares these binary eutectic melting curves, showing that while the Fe-FeH eutectic melting temperature is relatively low at pressures below ~120 GPa, as low as for the Fe-$Fe_3S$ system at 40 GPa, it is high, next to that of Fe-FeSi, at the ICB pressure of 330 GPa because of a large $dT/dP$ slope.

Here we discuss the effect of each light element on depressing the melting temperature of iron based on a combination of eutectic temperature and eutectic liquid composition in each binary Fe alloy system. The present experiments show not only the Fe-FeH eutectic temperature but also the possible range of eutectic liquid composition (Figure 5). The hydrogen content in liquid coexisting with the fcc phase gives the upper bound for that in the eutectic liquid. In addition, hydrogen concentration in the hcp phase found near the melting temperature also provides its lower bound (Table 2). These data suggest that the Fe-FeH eutectic liquid composition is about $FeH_{0.6}$ practically independent of pressure (Figure 5). It is more hydrogen-rich than the previous report of $FeH_{0.42}$ at ~40 GPa by Oka et al. (2022), possibly because their estimate based on an extension of a cotectic line in the Fe-O-H ternary system included some uncertainty. The d$T$/d$P$ slope of the Fe-FeH eutectic melting curve is comparable to those for the melting curves of pure Fe (Anzellini et al., 2013) and stoichiometric FeH endmembers (Tagawa, Helffrich et al., 2022) (Figure 6, 7), suggesting that the Fe-FeH eutectic liquid composition of $FeH_{0.6}$ changes little with increasing pressure to 330 GPa. If this is the case, the addition of 1 wt% H reduces the liquidus temperature of Fe by 1440 K at the ICB, when considering the eutectic temperature of 4700 K and the Fe melting point of 6230 K from Anzellini et al. (2013) (Table 3). Similarly, each binary eutectic liquid composition has been estimated at 330 GPa to be Fe containing 7.7 wt% S (Thompson et al., 2022), 8.0 wt% Si (Hasegawa et al., 2021), 18.4 wt% O (Sakai et al., 2022), and 3.5 wt% C (Mashino et al., 2019). Combining with their binary eutectic temperatures (Figure 7), the effects of depressing the liquidus temperature are found to be 290 K/wt% by S (Mori et al., 2017; Thompson et al., 2022), none by Si (Si has a negligible effect on the melting temperature depression according to Fischer et al., 2013), 130 K/wt% by O (Oka et al., 2019), and 720 K/wt% by C (Mashino et al., 2019) (Table 3).

In addition, we also compare such effect of liquidus temperature reduction considering the amount of each light element that reduces a certain percentage of liquid iron density. The density of liquid Fe is reduced by 1% by including 1.6 wt% S, 1.1 wt% Si, 1.0 wt% O, 0.78 wt% C, or 0.12 wt% H, practically independent of temperature, according to Umemoto & Hirose (2020) (see Table S1 in their paper). Such amounts of S, Si, O, C, and H depress the liquidus temperature of Fe at 330 GPa by 470 K, none, 130 K, 560 K, and 170 K, respectively (Table 3). It indicates that the core temperature is relatively low if C and/or S are important light elements largely contributing to the outer core density deficit, moderate when including H and/or O as major light elements, and high if Si is a predominant core light element, respectively. We discuss the liquid core temperature and composition in more detail below.

### 4.2. Estimates of ICB Temperature and Liquid Core Composition

The outer core liquid composition is required to explain the core density deficit, which is defined by the ICB temperature. And, the ICB temperature is the liquidus temperature of the outer core liquid at 330 GPa. The estimates of these two are not separate.

Regarding the ICB temperature, here we employ 6230 K for the melting temperature of pure Fe at 330 GPa (Anzellini et al., 2013), which is relatively high among those determined by static DAC experiments (see Fig. 4 in Ezenwa & Fei, 2023). The liquidus temperature of an Fe alloy was then estimated by simply adding together the effect of each light element (S, Si, O, C, and H) on depressing melting temperature discussed above. Indeed, such approximation of liquidus temperatures agrees reasonably well with earlier experimental results on ternary systems—Fe-C-O (Sakai et al., 2022), Fe-C-Si (Hasegawa et al.,2021; Miozzi et al., 2020), Fe-Si-H (Hikosaka et al., 2022), and Fe-O-H (Oka et al., 2022), in which liquid compositions including the carbon contents were quantified—performed at ~50 to ~200 GPa (the effect of melting temperature depression was evaluated at corresponding pressure). Figure 8 shows that the differences between the present estimate and the experimentally-obtained liquidus temperature are within ~10%.

When we estimate the possible outer core liquid composition, the S content is fixed to be 1.7 wt% based on cosmochemical and geochemical constraints (Dreibus & Palme, 1996; Hirose et al., 2021). We then widely explore the Si, O, C, and H contents in a liquid such that the overall light element concentrations explain the outer core density deficit under a certain ICB temperature (Umemoto & Hirose, 2020), which is calculated as the liquidus temperature of that liquid. Note that Umemoto & Hirose (2020)'s calculations showed negligible non-ideal mixing effect on volume for all combinations of liquid iron containing these light elements. We linearly interpolate and extrapolate the temperature dependence of their calculations.

The sets of possible outer core liquid composition and ICB temperature are illustrated in Figure 9, where the liquidus temperature of a given possible outer core liquid matches the ICB temperature. These results show that the ICB temperature strongly depends on the outer core carbon abundance (note that the sulfur content is fixed here, although it also strongly reduces the liquidus temperature). While it can be as high as 5720 K in the absence of carbon (Figure 9a), it might be 3730 K when 1 wt% C and 0.9 wt% H are included in the core (Figure 9b). Such ICB temperature of 3730–5720 K corresponds to the CMB temperature of 2740 to 4210 K when Grüneisen parameter is 1.5 (Vočadlo et al., 2003). Since the lowermost mantle is not extensively molten, it has been argued that the CMB temperature is lower than the solidus temperature of a pyrolitic mantle at 135 GPa, which may be as low as 3570 ± 200 K (Nomura et al., 2014) or 3430 ± 130 K (Kim et al., 2020) while earlier experimental studies reported 4180 ± 150 K (Fiquet et al., 2010) or 4150 ± 150 K (for a chondritic mantle by Andrault et al., 2011). Our experiments and modeling demonstrate that such low CMB temperature, less than 3430–3570 K, and the corresponding low ICB temperature of <4660–4860 K are indeed possible when the outer core is depleted in Si.

## 5. Conclusions

Fe forms a eutectic system with FeH above ~40 GPa (Tagawa, Helffrich et al., 2022), and here we determined the Fe-FeH eutectic melting temperature between 52 and 175 GPa using the H-poor hcp + H-rich fcc sample. Three independent melting criteria gave comparable eutectic temperatures at each pressure. Extrapolation of the eutectic melting curve shows that eutectic melting occurs at 4700 K and the ICB pressure of 330 GPa, which is lower than the eutectic temperature in Fe-FeSi but is higher than those in Fe-$Fe_3S$, Fe-FeO, and Fe-$Fe_7C_3$. The relatively high d$T$/d$P$ slope for the Fe-FeH eutectic melting curve is similar to those of the melting curves for Fe and stoichiometric FeH endmembers, suggesting that the binary eutectic liquid composition of $FeH_{\sim 0.6}$ found in this study might remain unchanged with increasing pressure to 330 GPa.

Based on such eutectic temperature and composition in the Fe-FeH system and those of other binary Fe alloy systems from literature, we found that the effect on liquidus temperature depression is large for C and S, moderate for H and O, and small for Si when considering the amount of each light element that causes a given reduction in liquid Fe density. Furthermore, we explored a possible liquid core composition in the Fe-1.7wt%S-Si-O-C-H system, which accounts for the outer core density deficit under an ICB temperature that matches the liquidus temperature of that liquid at 330 GPa. The results demonstrate that the ICB and overall core temperatures can be relatively low, even less than ~3500 K at the CMB when the outer core is poor in Si, which avoids extensive melting at the bottom of the mantle.

**Data Availability Statement**

Data for this research are found in Tables available at (Mita et al., 2024): Zenodo (https://doi.org/10.5281/zenodo.10947298).

**Acknowledgments**
We thank G. Helffrich and S. Fu for valuable discussion. Synchrotron XRD measurements were carried out at BL10XU, SPring-8 (proposals no. 2023A0181, 2023B1697, and 2023B1140). We thank K. Oka and S. Kawaguchi for their assistance in high-pressure gas loading. This work was supported by the JSPS grant 21H04968 to K.H.

**Table 1**

*Experimental Results*

| Run# | *P* (GPa) | Melting *T* (K) | | Melting *T* (K) from each criterion | | | | |
|---|---|---|---|---|---|---|---|---|
| | | Lower bound | Upper bound | Temperature plateau | Appearance of quench crystals | | Loss of one of solid phases | |
| | | | | | Lower bound | Upper bound | Lower bound | Upper bound |
| #1 | 52(3) | 1850(90) | 2080(100) | 2070–2080 | 1850 | 2070 | 1850 | 2070 |
| #2 | 62(3) | 2060(100) | 2100(110) | 2060–2100 | 2080 | 2100 | 2060 | 2080 |
| #3 | 116(6)<br>118(6) | 2710(140) | <br>3060(150) | —[a] | 2710 | 3060 | 2710 | 3060 |
| #4 | 148(15)<br>151(15) | 2590(130) | <br>3150(160) | —[a] | 2590 | 3150 | —[b] | |
| #5 | 159(8) | 3200(160) | | Not melted | Not melted | | Not melted | |
| #6 | 174(17)<br>175(18) | 3220(160) | <br>3400(170) | —[a] | 3220 | 3400 | —[c] | |

[a]Heating with higher laser power output was not performed after the appearance of quench crystals from liquid.

[b]Both hcp and fcc phases were present during heating.

[c]Only fcc phase was found before melting.

**Table 2**

*Observed Hydrogen Contents*

| Run# | Slightly above eutectic temperature | | | | Below eutectic temperature | | | |
|---|---|---|---|---|---|---|---|---|
| | *P* (GPa) | *T* before quench (K) | Liquidus phase | Liquid | *P* (GPa) | *T* (K) | hcp | fcc |
| #1 | 52(3) | 2070(100) | fcc, $x$=0.87(6) | $x$=0.71(5) | 52(3) | 1610(80) | $x$=0.51–0.67 | $x$=0.86–0.93 |
| #2 | 62(3) | 2150(110) | fcc, $x$=0.89(6) | $x$=0.78(5) | 61(3) | 1930(100) | $x$=0.41–0.65 | $x$=0.83–0.96 |
| #3 | 118(6) | 3060(150) | fcc, $x$=0.93(7) | $x$=0.74(5) | 116(6) | 2710(140) | $x$=0.49–0.61 | $x$=0.79–0.96 |
| #4 | 151(15) | 3150(160) | hcp, $x$=0.39(3) or fcc, $x$=0.95(7) | — | 148(15) | 2590(130) | $x$=0.22–0.46 | $x$=0.96 |
| #5 | Not melted | | | | 159(8) | 3200(160) | $x$=0.18–0.33 | $x$=0.93–0.99 |
| #6 | 175(18) | 3400(170) | fcc, $x$=1.05(7) | $x$=0.67(5) | 174(17) | 3220(160) | — [a] | $x$=0.96–1.05 |

[a]Only fcc phase was found.

**Table 3**

*Effect on Liquidus Temperature Depression*

| Light element | Eutectic *T* (K) at 330 GPa | Eutectic composition (wt%) | Liquidus *T* depression (K/wt%) | Amount to reduce liquid Fe density by 1% (wt%) | Liquidus *T* depression (K/1% density reduction) |
|---|---|---|---|---|---|
| S | 3990[a] | 7.7[a] | 290 | 1.6[h] | 470 |
| Si | 6700[b] | 8.0[e] | 0[g] | 1.1[h] | 0 |
| O | 3850[c] | 18.4[f] | 130 | 1.0[h] | 130 |
| C | 3710[d] | 3.5[d] | 720 | 0.78[h] | 560 |
| H | 4700 | 1.1 ($FeH_{0.6}$) | 1440 | 0.12[h] | 170 |

[a]Thompson et al. (2022), [b]Fischer et al. (2013), [c]Oka et al. (2019), [d]Mashino et al. (2019), [e]Hasegawa et al. (2021), [f]Sakai et al. (2022), [g]Assumed, [h]Umemoto & Hirose (2020)

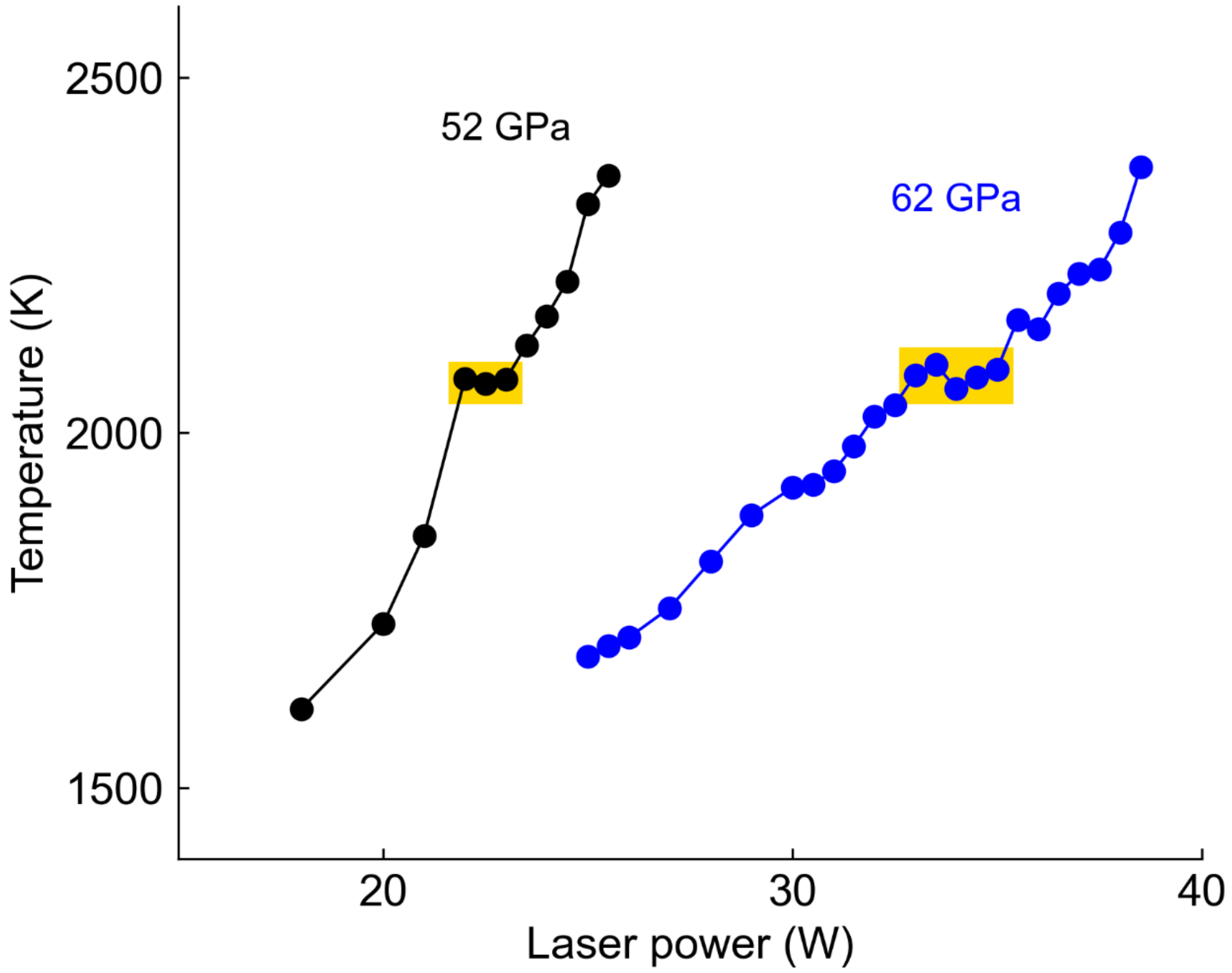

**Figure 1.** Laser power output vs. sample temperature relation. A temperature plateau (yellow part) may indicate melting temperature.

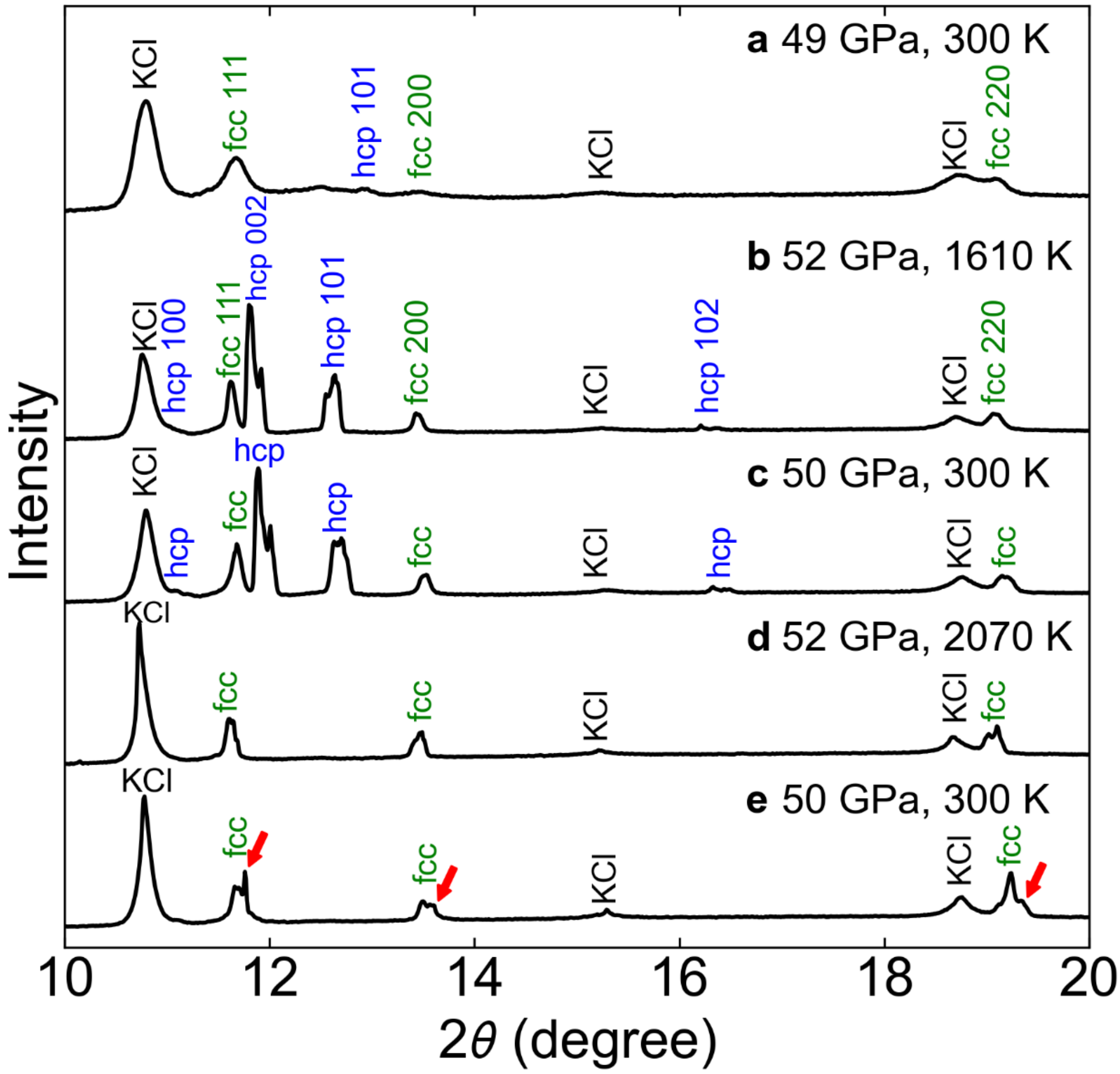

**Figure 2.** Changes in XRD patterns in run #1; (a) before heating, (b, c) during heating at 1610 K and after quenching to 300 K, and (d, e) during heating at 2070 K and after quenching. (b, c) Hydrogen-poor hcp $FeH_x$ ($x$ = 0.51–0.67) and hydrogen-rich fcc $FeH_x$ ($x$ = 0.86–0.93) coexisted at before melting. No new peaks appeared upon quenching temperature. (d, e) The hcp peaks disappeared upon heating, and quench crystals of fcc $FeH_{0.71}$ formed (red arrows) upon temperature quench, indicating the melting of a sample at 2070 K.

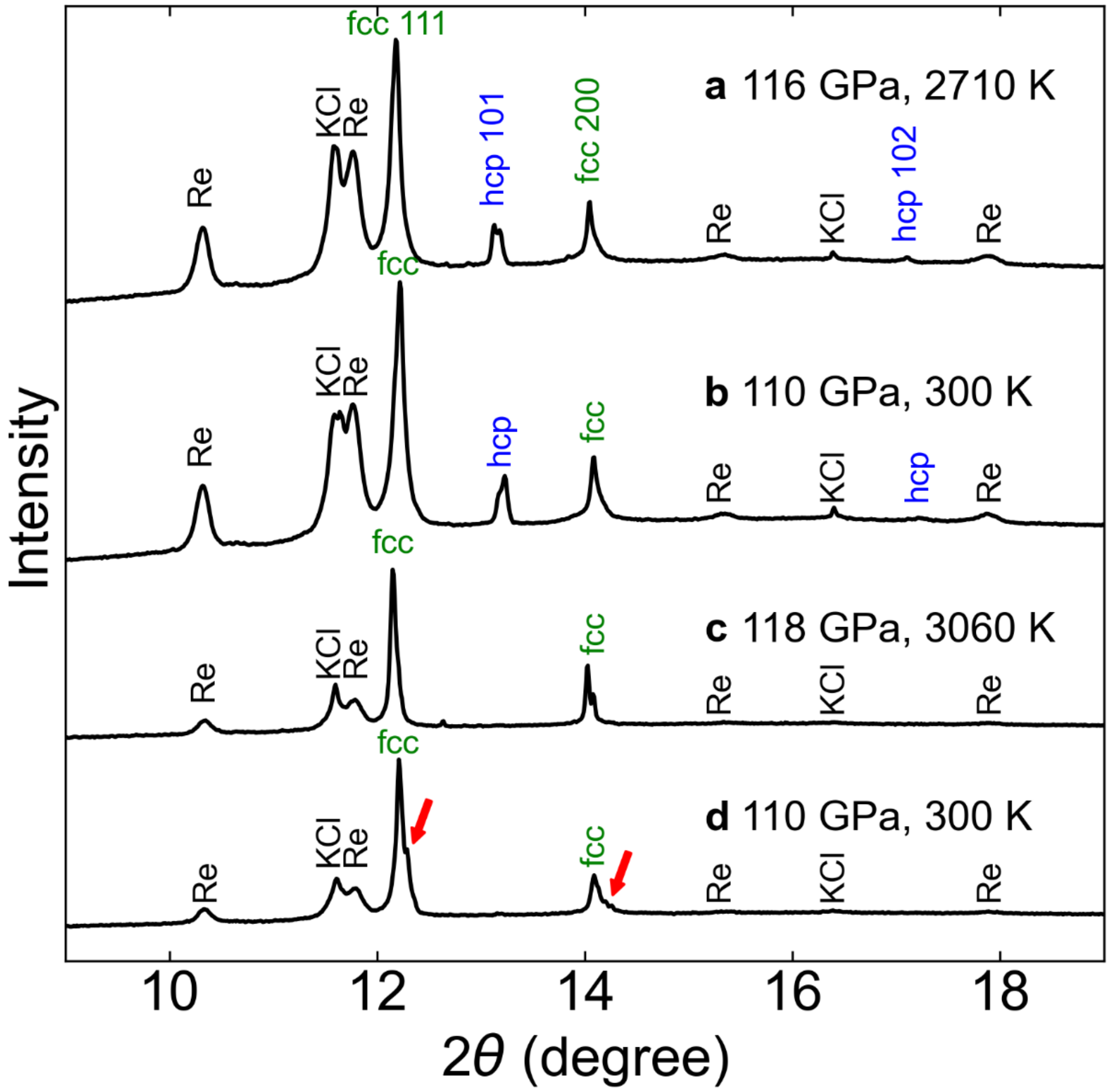

**Figure 3.** XRD patterns collected in run #3; (a, b) during heating to 2710 K and after quenching to 300 K, (c, d) during heating to 3060 K and after temperature quench. (a) Hydrogen-poor hcp $FeH_{0.49–0.61}$ and hydrogen-rich fcc $FeH_{0.79–0.96}$ coexisted at 2710 K. (b) No new peaks appeared upon quenching. (c) Hcp $FeH_x$ disappeared, indicating melting. (d) Fcc $FeH_{0.74}$ quench crystals appeared (red arrows), supporting the presence of liquid at 3060 K.

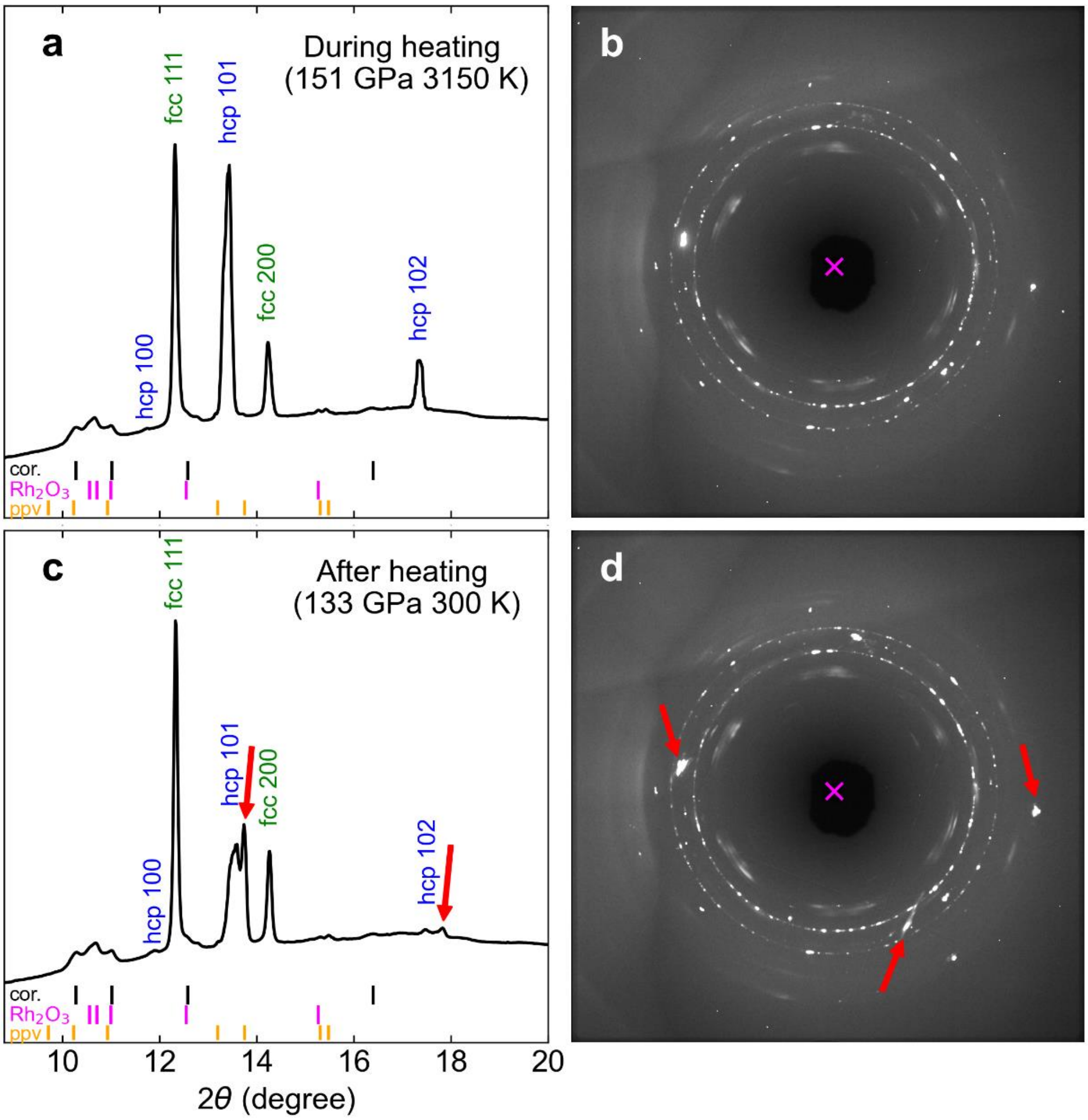

**Figure 4.** 1D and 2D XRD patterns obtained in run #4. Upon gradually decreasing temperature, the diffraction spots from hcp were stretched in the 2D image (marked by red arrows), suggesting the grain growth of the liquidus phase. The peak positions for corundum, $Rh_2O_3$-type, and post-perovskite-type $Al_2O_3$ are indicated by small bars in (a) and (c).

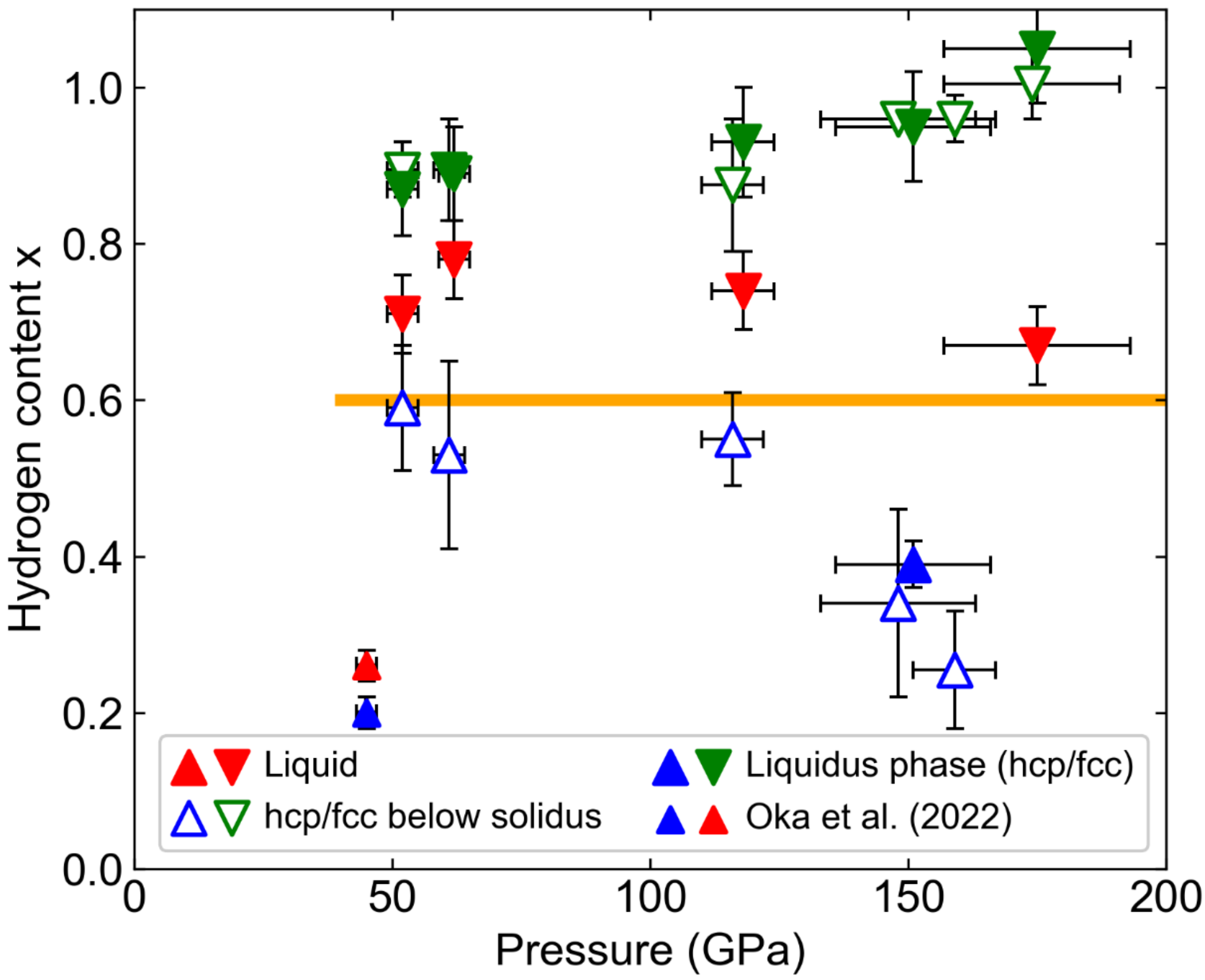

**Figure 5.** The hydrogen contents in subsolidus (open symbols) and liquidus (filled symbols) phases (hcp, blue; fcc, green) and quench crystals formed from liquid (red). Normal and inverted triangles indicate the lower and upper bounds for hydrogen concentration in the Fe-FeH eutectic liquid, respectively, which may be about $x = 0.6$ and almost independent of pressure (yellow line). See text for the pressure evolution of the eutectic liquid composition.

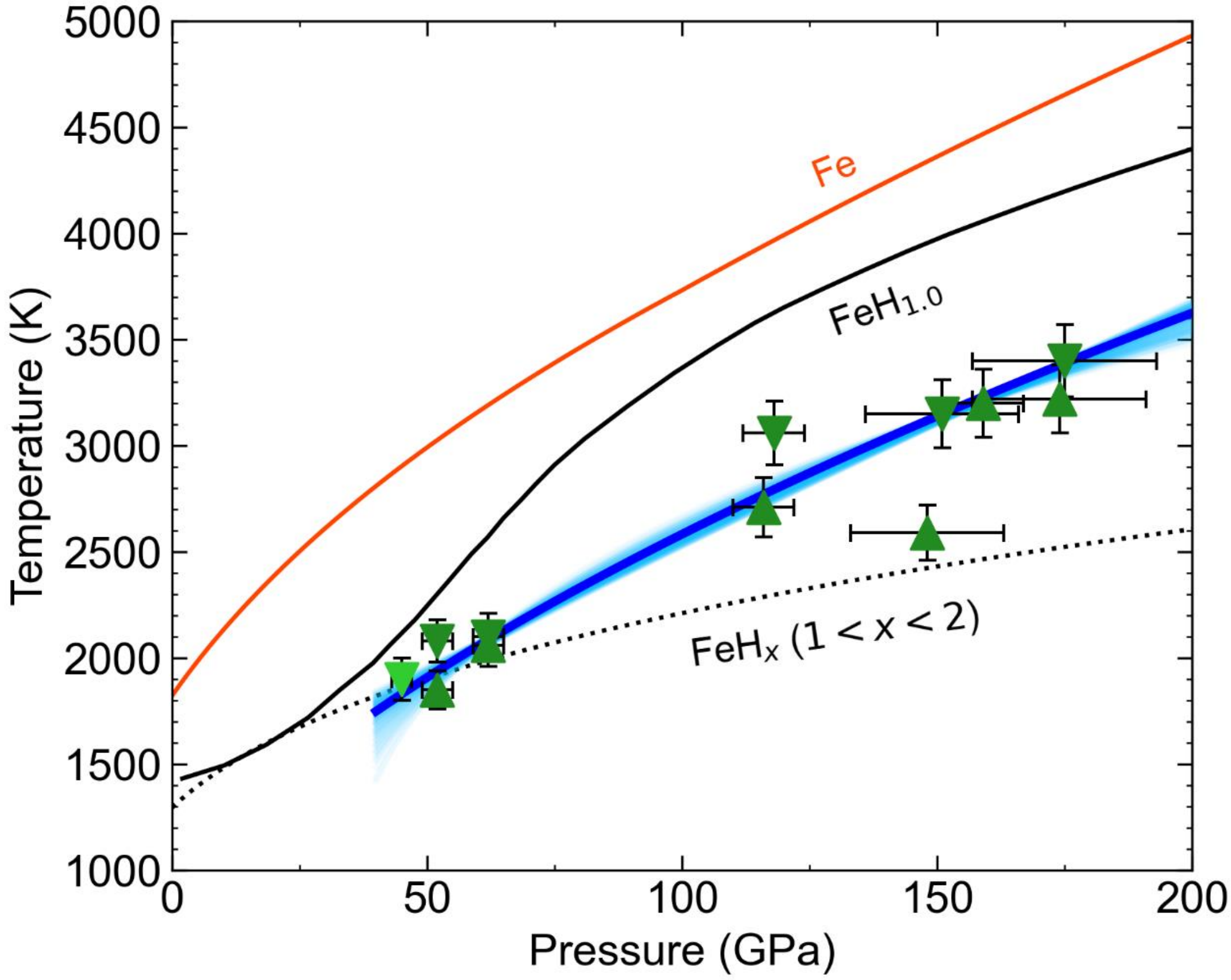

**Figure 6.** The Fe-FeH eutectic melting curve (blue line with light blue uncertainty band). Normal and inverted triangles show the lower and upper bounds for the eutectic melting temperature, respectively (dark green, this study; light green, Oka et al., 2022). Its temperature/pressure slope is similar to those of the melting curves of Fe (Anzellini et al., 2013) and stoichiometric FeH endmembers (Tagawa, Helffrich et al., 2022). The eutectic melting curve for $FeH_x$ ($1 < x < 2$) (Hirose et al., 2019) is also given by a dotted line.

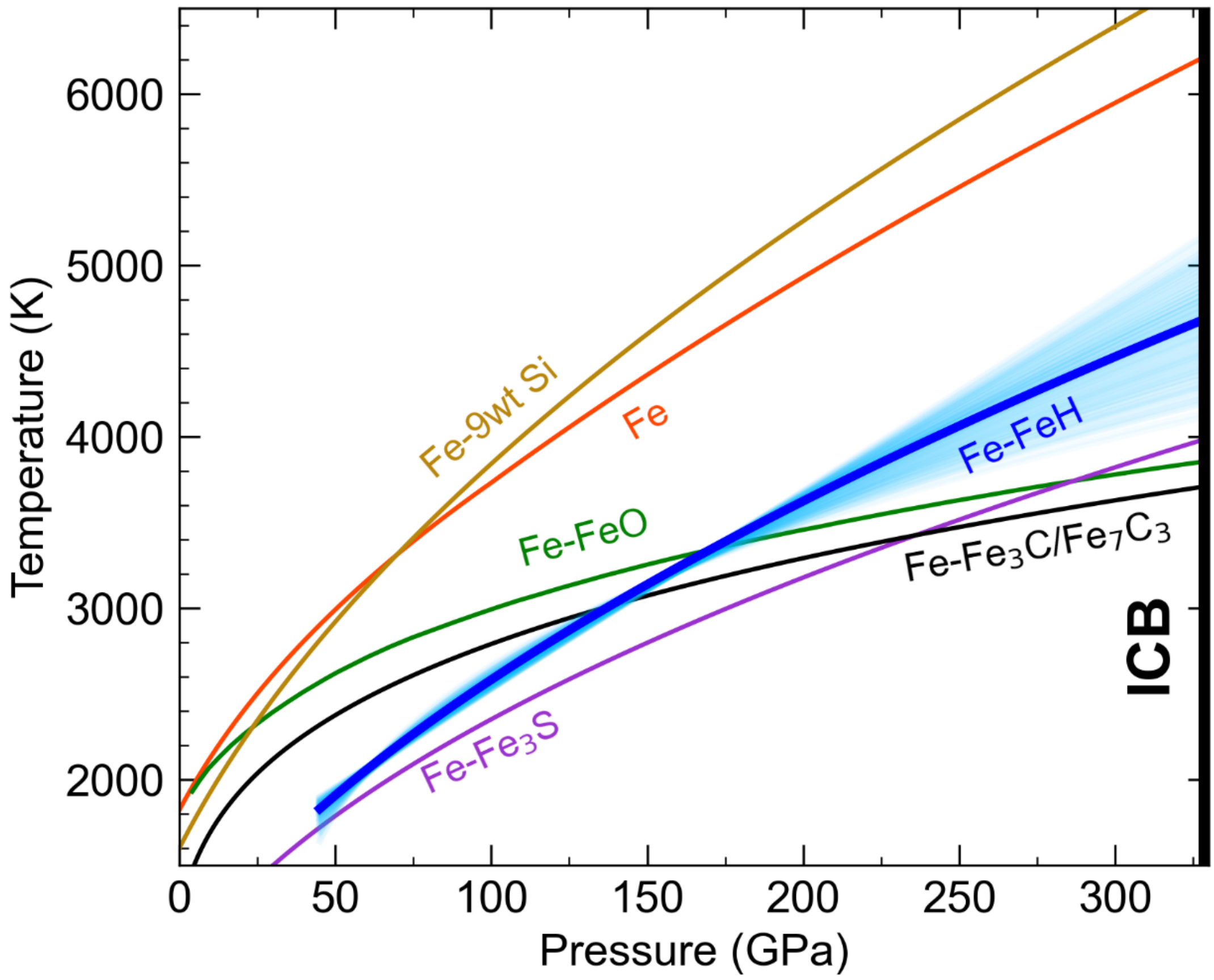

**Figure 7.** Comparison of the Fe-FeH eutectic melting curve with those for other binary Fe alloy systems; $Fe-Fe_3S$ (Mori et al., 2017), Fe-FeSi (Fe-9wt%Si) (Fischer et al., 2013), Fe-FeO (Oka et al., 2019), and $Fe-Fe_3C/Fe_7C_3$ (Mashino et al., 2019).

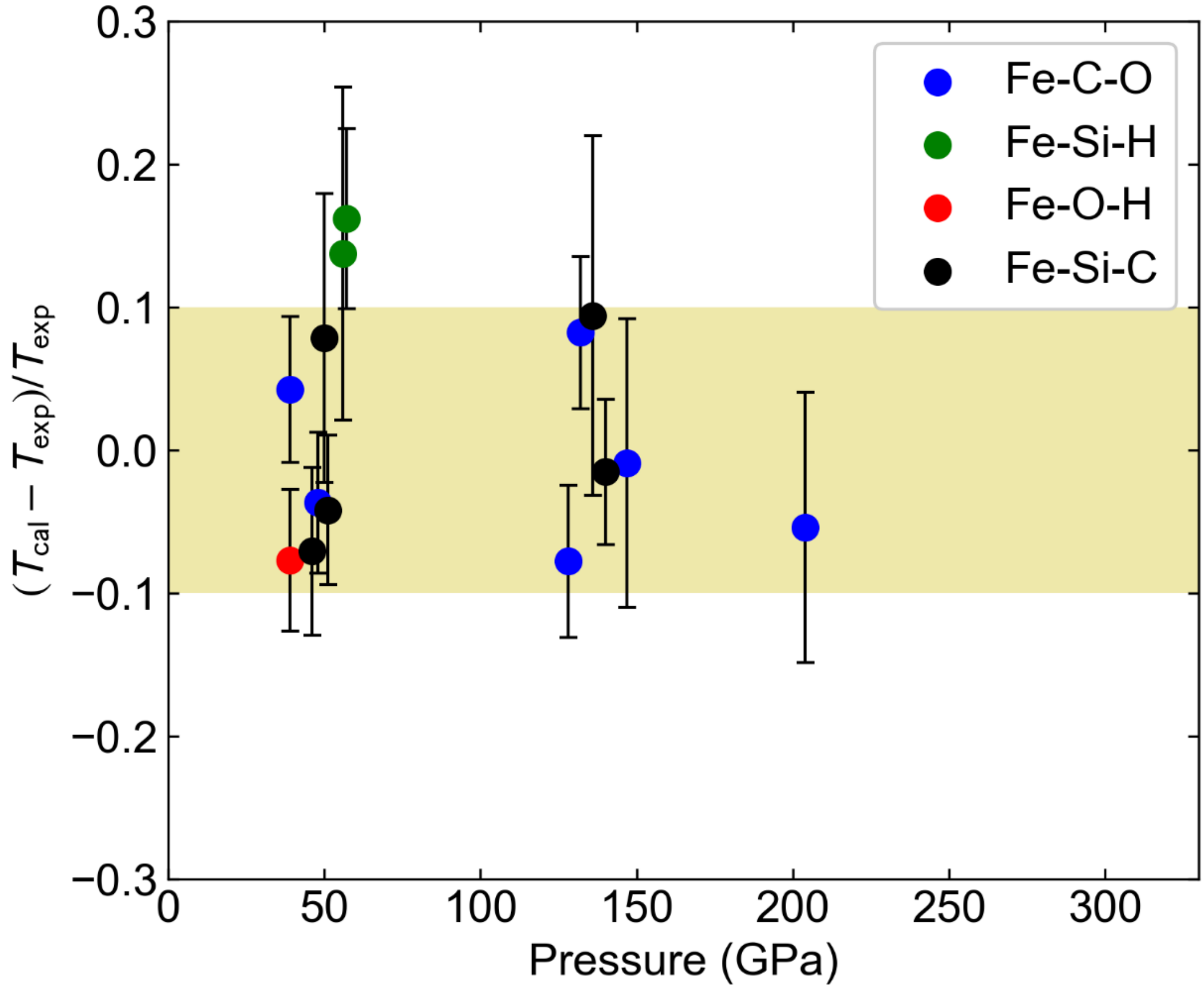

**Figure 8.** Differences in liquidus temperature between the present calculations and earlier experimental reports on ternary Fe alloys. Blue, Fe-C-O from Sakai et al. (2022); green, Fe-Si-H from Hikosaka et al. (2019); red, Fe-O-H from Oka et al. (2022); black, Fe-Si-C from Hasegawa et al. (2021) and Miozzi et al. (2020). We chose the data, for which Fe is liquidus phase and carbon concentration does not exceed the Fe-C eutectic liquid composition.

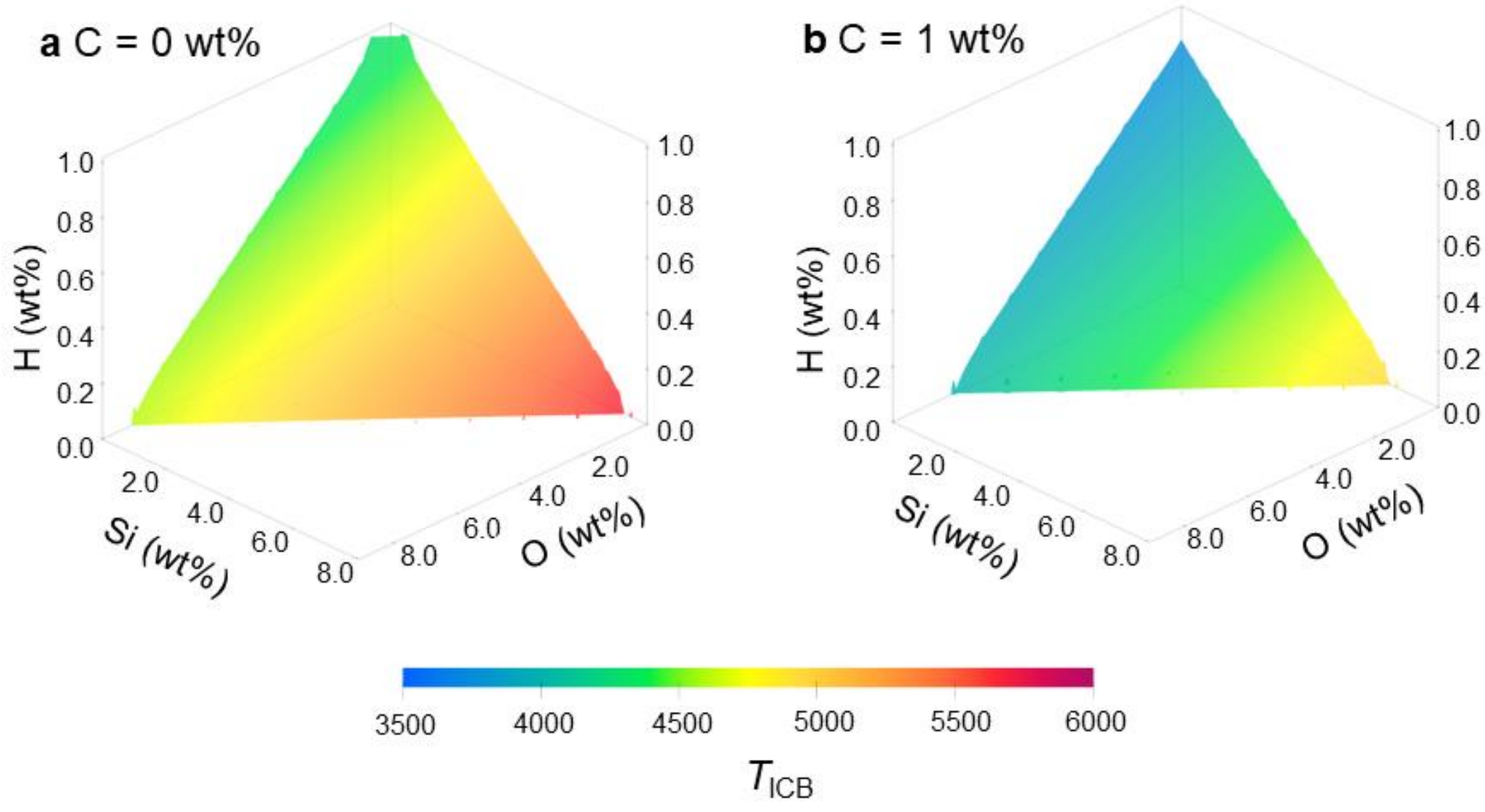

**Figure 9.** Possible liquid outer core compositions in Fe-1.7wt%S-Si-O-C-H with (a) 0 wt% and (b) 1 wt% C. Their liquidus temperatures calculated at 330 GPa (= ICB temperature) are shown by colors. These liquid compositions account for the outer core density deficit under the calculated ICB temperatures. Note that the density deficit with respect to pure Fe is larger for lower ICB temperature. See text for details.